# Soft-magnetic skyrmions induced by surface-state coupling in an intrinsic ferromagnetic topological insulator sandwich structure

Takuya Takashiro[1], Ryota Akiyama[1*], Ivan A. Kibirev[2], Andrey V. Matetskiy[2], Ryosuke Nakanishi[1], Shunsuke Sato[1], Takuro Fukasawa[3], Taisuke Sasaki[4], Haruko Toyama[1], Kota L. Hiwatari[1], Andrey V. Zotov[2], Alexander A. Saranin[2], Toru Hirahara[3] and Shuji Hasegawa[1]

[1]*Department of Physics, The University of Tokyo, Bunkyo, Toyko 113-0033, Japan*

[2]*Institute of Automation and Control Processes, 690041, Vladivostok, Russia*

[3]*Department of Physics, Tokyo Institute of Technology, Tokyo 152-8551, Japan*

[4] *National Institute for Materials Science (NIMS), 1-2-1 Sengen, Tsukuba, Ibaraki 305-0047, Japan*

**A magnetic skyrmion induced on a ferromagnetic topological insulator (TI) is a real-space manifestation of the chiral spin texture in the momentum space, and can be a carrier for information processing by manipulating it in tailored structures. Here, we fabricate a sandwich structure containing two layers of a self-assembled ferromagnetic septuple-layer TI, $Mn(Bi_{1-x}Sb_x)_2Te_4$ (MnBST), separated by**

---

* akiyama@surface.phys.s.u-tokyo.ac.jp

**quintuple layers of TI, $(Bi_{1-x}Sb_x)_2Te_3$ (BST), and observe skyrmions through the topological Hall effect in an intrinsic magnetic topological insulator for the first time. The thickness of BST spacer layer is crucial in controlling the coupling between the gapped topological surface states in the two MnBST layers to stabilize the skyrmion formation. The homogeneous, highly-ordered arrangement of the Mn atoms in the septuple-layer MnBST leads to a strong exchange interaction therein, which makes the skyrmions "soft magnetic". This would open an avenue towards a topologically robust rewritable magnetic memory.**

Real-space phenomena caused by non-trivial nature denoted by non-zero Chern numbers in topological insulators (TIs) have attracted much attention recently because of its novel Berry curvature physics such as quantum spin Hall effect. Once time-reversal symmetry (TRS) is broken by introducing ferromagnetic (FM) perturbation into a TI, gap-opening at the Dirac point and enhancement of the Berry curvature occur, resulting in e.g. quantum anomalous Hall effect[1-5] and the formation of skyrmions[6-13]. In particular, the correspondence between the real space and momentum space with respect to the chiral spin-texture induces a novel geometric effect, topological Hall effect (THE), which is one evidence for the non-trivial topological nature of the skyrmion systems as shown in Fig. 1a. Skyrmions have previously been reported in magnetically-doped TIs and ferromagnet/TI heterostructures[9-13] in addition to traditional trivial magnetic materials having the Dzyaloshinskii-Moriya interaction (DMI)[14-18]. They are observed transiently in the process of reversing the applied magnetic field because the strength of magnetic field for the magnetization reversal in the skyrmion cores is different from that in the matrix. Particularly, an interesting and unique point of skyrmions in FMTIs is that the Ruderman-Kittel-Kasuya-Yosida (RKKY) mechanism by the surface Dirac electrons is expected to mediate the DMI[6,7]. Thus understanding the interaction of magnetic moments with the Dirac surface states is essential to create, manipulate and annihilate skyrmions for application. The van der Waals material $MnBi_2Te_4$, which is a septuple layer (SL) where a Mn-Te layer is inserted in each $Bi_2Te_3$ quintuple layer (QL) as illustrated in Figs. 1b and c, is a model system to study the interplay between topology and magnetism and has been investigated extensively as an intrinsic self-assembled FMTI[5,19-25]. However, since the bulk conduction band crosses the Fermi level ($E_F$)[5,20,23,24,25], it is not an ideal system for the observation of the THE.

Therefore in this study, we focused on the Fermi-level-tunable FMTI, $Mn(Bi_{1-x}Sb_x)_2Te_4$ (MnBST) with varying $x$, which consists of SLs like $MnBi_2Te_4$[5,19-25] and $MnBi_2Se_4$[26] cases. In this system, the highly-ordered single Mn monolayers have both intra- and inter-layer long-range FM coupling, in contrast to the antiferromagnetic (AFM) nature of multi-layered MnBST[27]. The magnetization is also homogeneous compared to the randomly magnetically-doped FMTIs in the previous reports[3,4,9,10,12, 28 , 29]. By fabricating a MnBST/BST/MnBST (SL/QL/SL) sandwich structure and tuning the $E_\mathrm{F}$, we successfully observed skyrmion formation through detection of the THE in intrinsic magnetic topological insulators for the first time. Then precise control of the thickness of the BST spacer is found to be crucial to stabilize the skyrmions since the two chiral surface states of the MnBST need to have adequate coupling. This fact also rules out the possibility of the "mimic THE" discussed in magnetic impurity doped systems[29]. In addition, we found that the skyrmions in the present system have soft-magnetic properties compared to the doped magnetic topological insulator samples[10]. This is likely due to the homogeneous, well-ordered Mn arrangement since the exchange interaction within the Mn monolayers becomes nearly constant in terms of the sign and magnitude in the RKKY picture.

The samples were fabricated on a Si(111) substrate in ultra-high vacuum using the molecular beam epitaxy (MBE) method with stepwise growth of $(Bi_{1-x}Sb_x)_2Te_3$ (BST) and MnTe incorporation as in the previous work[20,25,26]. The sharp 1×1 pattern observed by reflection high-energy electron diffraction (RHEED) just after the growth of the top MnBST layer is shown in Fig. 1d, indicating high quality of the epitaxial and single-crystalline flat thin film. The atomic structure was analyzed by the high-angle annular dark field scanning transmission electron microscope (HAADF-STEM) imaging. Figure 1e shows a cross-sectional HAADF-STEM image recorded from the $[\bar{1}10]$ zone axis of the Si substrate. The additional top QL to a sandwich structure SL/QL/SL seems automatically assembled because of the local minimum of the formation free energy. Figure 1f displays the atomic-resolution HAADF-STEM image of the QL/SL/QL/SL stacking (left) and energy dispersive X-ray spectroscopy (EDX) map of Mn element (right) obtained from the same region of interest. In the EDX elemental map, the Mn element is dominantly incorporated in the middle of SLs, which indicates that a single Mn-Te layer is inserted in a BST-QL to form a self-assembled MnBST-SL (Fig. 1c) as explained before.

To observe the skyrmion-induced THE, the $E_F$ needs to be tuned within the bulk gap, that is, near the charge neutral point (CNP). For this purpose, Sb content $x$ was varied (0.20, 0.50, 0.55, 0.60) in $(Bi_{1-x}Sb_x)_2Te_3$ of all MnBST and BST layers. Fig. 2a shows magnetic field dependences of the transverse (Hall) resistivity $\rho_{yx}$ of samples with different $x$ at 2 K. They all show hysteretic loops indicating AHE with the most pronounced effect at $x = 0.55$. The results mean that not the AFM but FM order is induced in every sample.

To infer the $E_F$ position, the carrier type and density were estimated as shown in Fig. 2b from the component of the ordinary Hall effect. It should be noted that the $\rho_{yx}$ in a FM material is typically described as $\rho_{yx} = R_0\mu_0 H + \rho_A + \rho_T$ ; the first term denotes the ordinary Hall resistivity where $R_0$ and $\mu_0 H$ are the Hall coefficient and the applied surface-normal magnetic flux density, respectively[28]. The second and third terms represent the anomalous Hall resistivity which is in proportional to the magnetization and the topological Hall resistivity due to skyrmions if any, respectively. As shown in Fig. 2b, the carrier density and type which are evaluated from $R_0$ change with *x* and the $E_F$ can be tuned near the CNP by controlling *x* to be 0.50 - 0.55. In MnBi$_2$Te$_4$, it was reported that increasing the Sb content substituted for Bi leads to the change of the carrier density and the carrier type, from *n*- to *p*-types due to the carrier compensation by hole-doping from Sb atoms, resulting in enabling the $E_F$ control through tuning Bi/Sb ratio[27,30].

In Fig. 2a, hump-like features in the curves were observed as indicated by hollow arrows, which is one of typical characteristics of THE, as discussed later in detail. Here, we introduce parameters to analyze magnitudes of AHE and THE quantitatively: because no skyrmions emerge under the saturation field (such as $\mu_0 H$ = 2.5 T), the topological Hall resistivity $\rho_T$ does not contribute to $\rho_{yx}$, so that the anomalous Hall resistivity at the saturation magnetization field $\rho_A^{sat}$ is represented by $\rho_A^{sat} = \rho_{yx} - R_0\mu_0 H$ at 2.5 T. On one hand, $\rho_T$ contributes only near the hump, so that $|\rho_T^{peak}| = |\rho_{A+T}^{peak}| - |\rho_A^{sat}|$ where $\rho_{A+T}^{peak}$ is $\rho_A + \rho_T$ at the peak of the hump. In Figs. 2c and d, it is shown that the sample with $x = 0.55$ has the maximal $|\rho_{yx}|$ at $H = 0$ among the samples, and magnitudes of both the AHE and THE components (namely $|\rho_A^{sat}|$ and $|\rho_T^{peak}|$) are enhanced as $E_F$ becomes closer to the CNP. This enhancement does not originate from the extrinsic scattering on magnetic impurities, but from the intrinsic magnetic scattering caused by the Berry curvature of the energy band with the TRS-broken gap around the CNP[31]. Also in the previous paper, C. Liu *et al.* [10] has reported that the THE as well as the AHE are enhanced when $E_F$ lies crossing the chiral spin texture band located in the bulk gap because the contribution from the magnetized band having a spin chirality increases as schematically shown in Fig. 2e.

To confirm the skyrmion nature of the observed Hall resistance anomalies, we focus on the $x = 0.55$ sample whose $E_F$ is closest to the CNP. Figure 3a shows $\rho_{A+T}$ as a function of the applied magnetic field $\mu_0 H$ at temperatures 0.5 K – 8 K. Below 5 K with lowering the temperature, the THE humps associated with the skyrmions as well as the AHE loops become more pronounced. Figure 3b displays magnetic field dependences of magnetoresistivity (MR) $\rho_{xx}(H) - \rho_{xx}(0)$ at temperatures of 0.5 K – 20 K. Below around 7 K, negative MR is seen at weak magnetic field which is defined by the subtraction of the minimum of $\rho_{xx}$ from $\rho_{xx}$ at 0 T, and it almost disappears at $T^* \sim 8$ K and above as shown by Fig. 3c. The MR turns to positive above $T^*$ as described in detail in Supplementary Section V.

Temperature dependences of AHE and THE, $|\rho_A^{sat}|$ and $|\rho_T^{peak}|$, are plotted in Fig. 3d. $|\rho_A^{sat}|$ is enhanced with decreasing temperature below the Curie temperature ($T_C$ = 22 K) which is estimated by the Arrott plots (see Supplementary Section VI). $|\rho_T^{peak}|$ also becomes larger below around $T^*$, which is consistent with the behavior of the MR shown in Fig. 3c. For comparison of the components of AHE and THE more quantitatively, we define the coercivity of $\rho_{A+T}$ as $H_C$, and $H_T$ corresponding to the peak position of the hump. These values at each temperature are plotted in Fig. 3e. The PM-to-FM transition occurs at 22 K corresponding to $T_C$. On the other hand, below $T^*$, skyrmions start to appear under low magnetic fields less than ~ $H_T$, while the whole region changes to the FM order under magnetic fields higher than $H_C$. Figure 3e depicts the magnetic phase diagram where "PM" and "Sk" are paramagnetic and skyrmion states, respectively.

Figures 4a and b show the magnetic field dependences of $\rho_{A+T}$ and $\rho_{xx}$, respectively. The magnetic fields at the hump-like feature in Fig. 4a correspond to those of the MR peaks in Fig. 4b at 0.5 K as indicated by hollow arrows. This indicates that the carrier scattering due to the emerging magnetic field induced by skyrmions occurs most prominently at the applied magnetic field where the density of skyrmions is at the maximum, resulting in the MR peaks.

To reveal the coupling of the chiral states between the top and bottom FM SLs, we fabricated samples having the BST spacer of different thickness, namely, sandwich structures of SL-MnBST/$N$-QL-BST/SL-MnBST with $N$ = 0, 1, 2. The AHE hysteresis curves were observed in all samples at 0.5 K, and the curve of the sample with $N$ = 1 shows the largest as shown in Supplementary Section VII. The THE component, $\rho_\text{T}$ ($\rho_\text{A+T} - \rho_\text{A}^\text{sat}$) of the three samples at 0.5 K is shown in Fig. 4c as a function of applied magnetic field (sweeping from the negative magnetic field). The values of THE resistivity $\rho_\text{T}$ at the hump are 0, 145 and 26.1 Ω for the samples of $N$ = 0, 1, 2, respectively; $\rho_\text{T}$ is the largest for $N$ = 1, while the samples with $N$ = 2 and 0 have weaker or negligible THE. The 1QL spacer looks the best for producing skyrmions. This result is interpreted as follows. The DMI at the surface/interface of the magnetic TI is essential to form skyrmions, and the density and the size of skyrmions depend on the magnitude of DMI. Generally, on the other hand, since the spin chirality in the surface band is opposite between the top and bottom surfaces and the hybridization between the two surface states would depend on the spacer thickness[9,10], the surface-state-mediated DMI would also change with the spacer thickness. When the top and bottom SLs are so close and strongly interfere to each other, the DMI is significantly weakened by strong hybridization, resulting in opening a hybridization gap larger than the magnetic gap in each SL, and thus skyrmions are not likely to form. This is the case for the sample of $N = 0$. For the sample with $N = 2$, on one hand, the top and bottom SLs have negligible hybridization to be decoupled from each other and have the opposite spin chirality separately. Hence, the surface-state-mediated DMI has the opposite sign in the two surface states, which makes skyrmions frustrating and unstable. The sample with $N = 1$ would have intermediate hybridization between the top and bottom SLs to have the same spin chirality, leading to the same sign of their DMI, resulting in skyrmions formed stably (Fig. 4d). This significant dependence on the spacer thickness strongly indicates that the DMI is sensitive and crucial to the degree of the coupling between the topological surface states of the top and bottom SLs.

In addition to these experimental results of THE indicating skyrmions, we calculated the energetic stability of the skyrmion using the Hamiltonian of 3D tight-biding lattice model that includes the magnetic interaction and potential gradient along thickness direction for the BST/MnBST/BST/MnBST structure (the detail method is shown in Methods and Supplementary Section VIII)[9,10]. Here, we assume that the band structures near $E_F$ are almost unmodulated from the pristine TI's one by forming the magnetic SL, which have been shown by the previous calculation[22]. The sample with $x$ = 0.55 was $p$-type with the $E_F$ near the CNP and thus we simulated the model assuming that the $E_F$ lies slightly below the edge of the magnetic gap in the massive Dirac state (inset of Fig. 4e). Three typical domain walls in skyrmions are assumed; the Bloch, Néel 1 and Néel 2-types. Figure 4e shows the formation energy of a skyrmion relative to that of the spin-collinear FM for each domain case where $R$ indicates the skyrmion radius. When $R = 0$ the energy is minimum, meaning the FM state. The energies at $R \neq 0$ in the Bloch and Néel 1 cases are significantly higher than that at $R = 0$, indicating that the FM state is most stable. However, the Néel 2-type state shows the negative energy at $R \neq 0$, resulting in the stable ground state of skyrmions. This calculation indicates that the observed THE in our MnBST/BST/MnBST is derived by the Néel 2-type skyrmions, which also have been reported in the previous FMTI thin film systems[9,10].

Recently, K. M. Fijalkowski *et al.* have indicated that the characteristic hump-like curves observed in a magnetically-doped TI/nonmagnetic TI heterostructure can be induced by the combined signals of opposite-sign AHEs, which result from two FM orders in the bulk and at the surface, not by the skyrmion-induced THE[29]. The hump-like curves in our system, however, cannot be explained using the above model even though there are two different magnetic Mn layers in the sandwich structure. The detailed discussion is described in Supplementary Section IX.

As shown in Fig. 5, it is interesting to point out that the shape of our THE curves is apparently different from that of the Mn-doped FMTI thin films although the magnetic atom, Mn, is the same as our system. In the reported Mn-doped case[10], the sign of THE is negative (Fig. 5a), while in our case it is positive (Fig. 5b). In general, $\rho_T$ is proportional to the Hall coefficient $R_0$, and the sign of $\rho_T$ depends on both the sign of $R_0$ and the direction of emergent magnetic field $H_{em}$ which is produced by the skyrmions (Fig. 1a) and is in proportional to the magnitude of the hump-like peak[9,18]. In the cases of Figs. 5a and b, both systems are $p$-type, and thus the sign of $R_0$ is the same, which indicates that the sign of $\rho_T$ is determined by $H_{em}$.

The shape of $\rho_{A+T}$ curve, which can be decomposed to AHE and THE curves as schematically shown in Fig. 5, is related with the sign of THE, and is explained by the difference of the applied magnetic field necessary for magnetization reversal between in the matrix and at the cores of skyrmions. In the previously reported Mn-doped $Bi_2Te_3$ thin films[10], as shown by a red curve in Fig. 5a, with increasing the magnetic field from the negative, the magnetization of the matrix starts to reverse first from negative to positive before $H_C'$, remaining skyrmions having negative $H_{em}$. When the magnetic field exceeds $H_T$ ($H_C' < H_T$), skyrmions are wiped out so that the film is in FM state (positive magnetization). This is because under the strong magnetic field enough for saturating the whole magnetization, the Zeeman energy defeats the DMI, resulting in disappearance of skyrmions. In contrast, in our MnBST-SLs system the process of the magnetization reversal is opposite; as shown by a red curve in Fig. 5b, the magnetization of skyrmions starts to reverse first ($H_{em}$: from negative to positive) before $H_C'$ ($H_C' > H_T$), and then the magnetization in the matrix reverses by the larger magnetic field. This difference in the magnetization reversal corresponds to the opposite sign of $H_{em}$ compared to that of the Mn-doped $Bi_2Te_3$ films case. In addition, $H_T$ is much smaller in our system (~ 0.1 T at 0.5 K) than that in the previous system (~ 1 T at 1.5 K)[10]. Intriguingly, the reason of the opposite sign of $H_{em}$ and the smaller $H_T$ could be attributed to the kinds of magnetic interaction between nearest-neighbor Mn atoms. In TRS-broken FMTIs, it has been explained and demonstrated that the Dirac electrons at the surface state mediate the interaction between magnetic moments[32]. Therefore, the FM exchange interaction is based on the RKKY mechanism in addition to DMI which is induced by the strong spin-orbit coupling and the inversion symmetry breaking. This RKKY interaction, which strongly depends on the distance between magnetic atoms, determines $H_T$; the FM interaction $|J|$ is larger, the $H_T$ is smaller[33]. This is because in our case, the core-magnetization of skyrmions more easily aligns to the external magnetic field due to the large |$J$|, and thus skyrmions are generated at the lower magnetic field. In Mn-doped FMTI films, the nearest Mn-Mn distance (averagely ~ 0.73 nm)[10,34] is larger than that in our self-assembled MnBST-SL (~ 0.43 nm)[19]. In our case, the Mn atoms are densely ordered in the two-dimensional Mn single layer in SL, and thus |$J$| is expected to be larger than that of the Mn-doped FMTI films. The important point is that the sign of THE directly reflects the degree of order of Mn atoms and our system is found to have a strong exchange interaction between Mn atoms.

In conclusion, we observed the skyrmion-induced THE as well as the AHE in sandwich structures consisting of two layers of the self-assembled FMTI MnBST-SL separated by a non-magnetic BST layer. The $E_F$ position was tuned near the CNP by optimizing Sb content ($x$ = 0.55) to enhance the THE. It was clearly shown that the AFM coupling between the two SLs was suppressed by our designed sandwich structure. When no spacer layer between the two MnBST layers is inserted ($N$ = 0), the hybridization gap becomes dominant compared with the magnetic gap and topological states disappear. Meanwhile, when the BST spacer is two QL thick ($N$ = 2), the top and bottom MnBST layers are decoupled from each other, which leads to the electrons at the $E_F$ have the opposite spin chirality for the two MnBST layers on different surfaces that make skyrmions unstable. For the sandwich structure with $N$ = 1, the two surface states of MnBST-SLs having the same spin chirality couple with each other to stabilize skyrmions producing the largest THE. This indicates that the skyrmion state is controllable by tuning the degree of the coupling of surface states[35]. Moreover, compared to Mn-doped FMTI thin films, our sandwich structures with self-assembled FMTI have a denser arrangement of the Mn atoms, resulting in stronger exchange FM interaction between Mn atoms in the layer, and thus skyrmions can be created and annihilated at weaker magnetic field[10]. In other words, the magnetic field of magnetization reversal in skyrmions becomes smaller, leading to the different shape of the THE curves compared with the previously reported THE shapes. This topologically robust "soft magnetic" properties of skyrmions in our system would shed a light on the path for the topologically robust, but easy to be rewritten magnetic memory.

**Methods**

**Sample fabrication**

All samples were grown by MBE in an ultrahigh vacuum (UHV) chamber of $\sim 1\times 10^{-10}$ Torr equipped with a RHEED system. After preparing a clean Si(111)-7×7 surface on a $n$-type substrate (4 mm × 25 mm in size) by cycles of resistive heat treatment, one quintuple layer (QL) $(Bi_{1-x}Sb_x)_2Te_3$ (BST) was grown with co-evaporation of Bi and Sb on the substrate at ~ 200 °C in a Te-rich condition. Then, we grew one septuple layer (SL)

Mn($Bi_{1-x}Sb_x$)$_2$Te$_4$ (MnBST) by depositing Mn on 1QL BST at ~ 270 °C in a Te-rich condition. By using such procedures, we fabricated a smooth and epitaxial stacking structure of BST/MnBST/BST/MnBST on the substrate. Finally, to avoid possible oxidization, we deposited Al on top at room temperature to form a ~ 2-nm-thick $Al_2O_3$ cap layer when it is taken out in air.

**Structure analysis by STEM**

After depositing Pt on top as a protection layer, the samples were prepared into the electron transparent specimens for STEM observations by the standard lift-out technique using the FEI Helios G4-UX dual-beam system. Probe aberration corrected STEM (FEI TitanG2 80-200 microscope) was used. Chemical compositions were analyzed by energy-dispersive X-ray spectroscopy (EDX).

**Transport measurements**

All electrical transport measurements were performed with physical property measurement system (PPMS) by Quantum Design Inc. after Au wires were bonded to samples with the silver paste for the conventional 6-terminal methods. In the low-temperature measurements (< 2 K), the $^3$He option of the PPMS was used. All results of the Hall effect were obtained after the anti-symmetrized treatment as a function of the magnetic field. Detailed explanation can be found in Supplementary Information.

**Calculation of skyrmion stability**

The energetic stability of skyrmion configurations in the SL/QL/SL sandwich structure was calculated by the tight-binding lattice model for a 3D TI using the effective Hamiltonian in the continuum approximation[9,10],

$$H_0 = t\sum_{\alpha=x,y,z} \sin k_\alpha \sigma_\alpha \tau_x + \left[m + 2D_1 \sum_{\alpha=x,y,z}(1-\cos k_\alpha)\right]\tau_z. \quad (1)$$

We consider an exchange between electrons and magnetic moments as well as a surface potential difference in a magnetic TI phase ($U$) term,

$$H' = -(J_0 - J_3\tau_z)\mathbf{n}\cdot\boldsymbol{\sigma} + \left(-U + \frac{2zU}{N_z-1}\right) \quad (2)$$

where $\sigma$ and $\tau$ are the Pauli matrices for the spin and orbital degrees of freedom, respectively. $J_0$ and $J_3$ account for the electron-hole symmetric and asymmetric parts of exchange coupling[8]. In our calculation, we take $N_x = N_y = 16$ sites and $N_z = 4$ layers, and impose the periodic boundary condition in both $x$ and $y$ directions. Here, we assume that the band structures in the vicinity of the $E_F$ do not change by forming the magnetic SL from the pristine topological insulator's one, which have been shown by the first-principles calculation[22]. The parameters are set as $D_1 = 1$, $D_2 = 0.2$, $t = 1$, $m = -1$, which realizes the strong TI phase such as BST. We introduced the potential asymmetry between BST and MnBST by setting $U = 0$ for BST and $U = 0.05$ for MnBST, and the exchange couplings by setting $J_0 = J_3 = 0$ for BST and $J_0 = 0.1$ and $J_3 = 0.062$ for MnBST referring to the previously reported calculations[9,10]. The relationship between the skyrmion stability and $J_3/J_0$ is shown in Fig. S10 of Supplementary Information.

Here, we can expect three types of skyrmion configuration $\boldsymbol{n}(\boldsymbol{r})$, *i.e.*, the Bloch, Néel 1, and Néel 2 types, which are expressed generally as

$$\begin{cases} n_x = \sin\left[\pi(1-\frac{r}{R})\right]\cos(\theta+\phi) \\ n_y = \sin\left[\pi\left(1-\frac{r}{R}\right)\right]\sin(\theta+\phi) \\ \quad n_z = \cos\left[\pi(1-\frac{r}{R})\right] \end{cases} \tag{3}$$

for $r < R$. In the case of $r \geq R$, we take $n_x = n_y = 0$ and $n_z = 1$. $R$ is the skyrmion radius set as the variational parameter and $\phi$ determines the magnetic helicity or the skyrmion type, that is, $\phi = \pi/2, 0$ and $\pi$ corresponding to the Bloch, Néel 1 and Néel 2-types, respectively. Note that $R = 0$ corresponds the ferromagnetic phase. We calculated the total energy as a function of $R$ considering the reference value at $R = 0$.

**Acknowledgement**

We thank N. Nagaosa and M. Ezawa in the University of Tokyo for fruitful discussion on the mechanism of emergence of skyrmions. Electrical measurements were partially performed with using facilities set in H. Takagi group in Department of physics and in Cryogenic Research Center, the University of Tokyo. We appreciate cooperation by S. Ichinokura in Tokyo Institute of Technology for ARPES experiments and by K. Suzuki in NIMS for technical preparation of HAADF-STEM measurements. We are grateful to K. Hono in NIMS for discussing results and supervising of HAADF-STEM measurements. This research was partly supported by JSPS KAKENHI Grant Nos. 20H02616, 20H00342, 18H03877 and 17H06152 and the Russian Foundation for Basic Researches under Grant No. 19-02-00549.

**Author contributions**

TT fabricated samples with help of RA, RN and KLH. TT planned and performed electrical transport measurements with advised by RA and SH who conceive and design experiments. TT, HT, IAK, and AVM conducted ARPES measurements under the supervision of AVZ and AAS. SS numerically calculated the stability of skyrmion types. TT, KLH and TF performed complemental ARPES measurements on the advice of RA and TH. TT, RA and TS carried out STEM observations. All authors take part in discussing experimental results and interpretation. TT, RA, TH and SH wrote the manuscript with contributions from all the authors. SH supervised the whole project.

**Competing interests**

The authors declare no competing interests.

**Figures and Figure Captions**

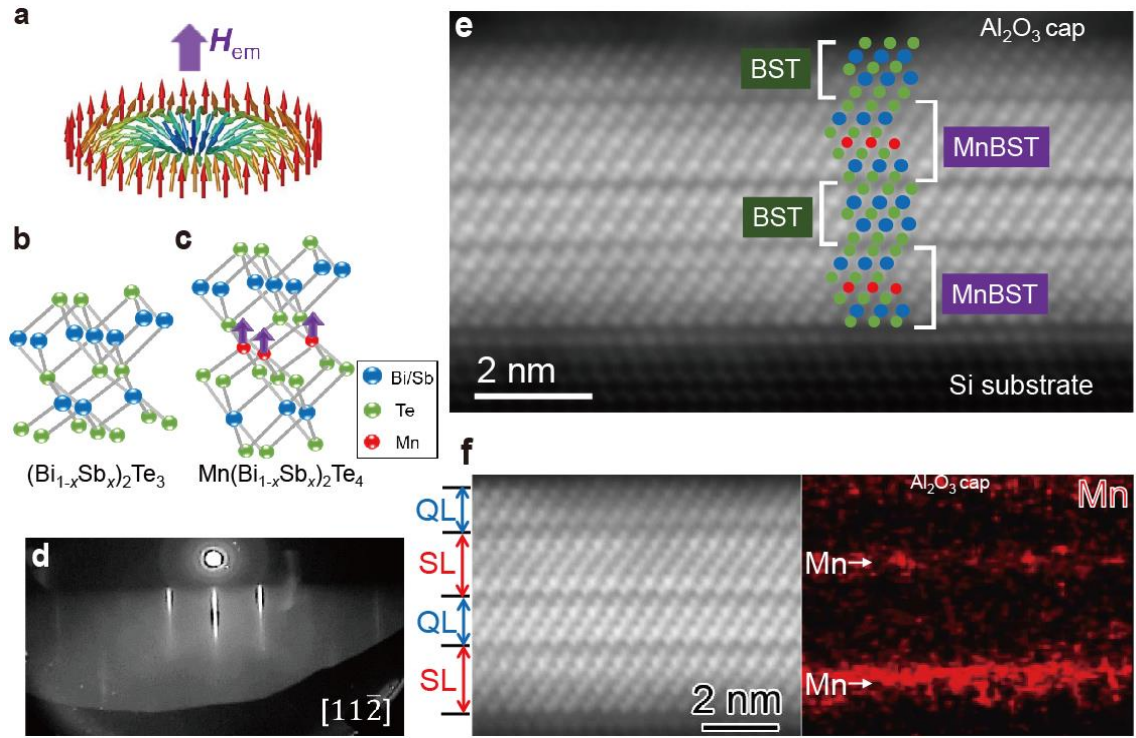

**Figure 1 | Properties of a skyrmion and the crystal structure in Mn(Bi$_{1-x}$Sb$_x$)$_2$Te$_4$ (MnBST).** **a,** Schematic picture of magnetization of a skyrmion. The magnetic texture has an emergent magnetic field represented by $H_{em}$. **b, c,** Schematics of BST quintuple layer (QL) and MnBST septuple layer (SL) with an intercalated ferromagnetic Mn layer. **d,** RHEED pattern observed from [11$\bar{2}$] of the Si(111) substrate just after the growth of the top MnBST layer. **e,** Cross-sectional HAADF-STEM image of the sample with $x = 0.55$, showing the stacking structure consisting of QL and SL. **f,** Atomic-resolution HAADF-STEM image (left) and EDX elemental map of Mn (right). Except for the Al$_2$O$_3$, layer, the Mn distribution in SLs indicates that a single Mn-Te layer is intercalated into 1 QL BST, resulting in the self-assembled MnBST SL.

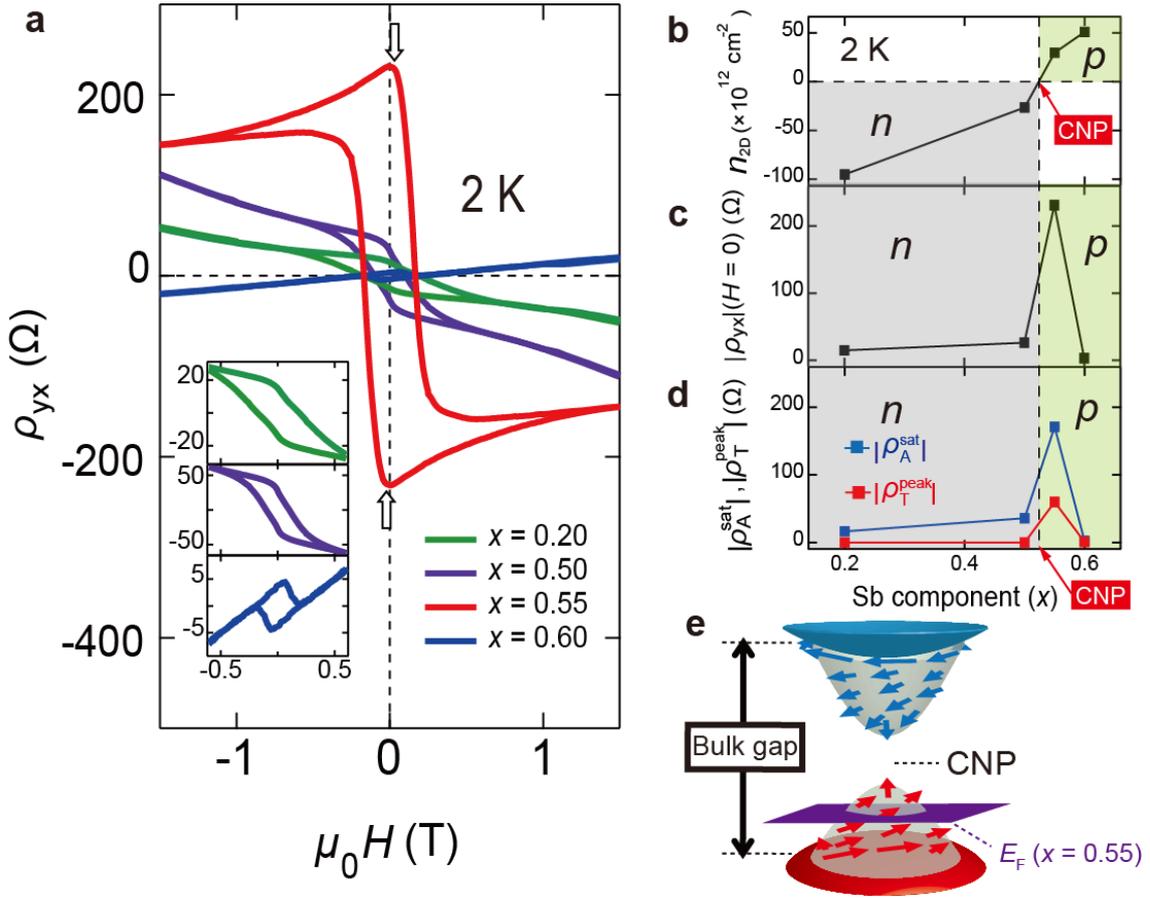

**Figure 2 | Hall measurements in MnBST (1 SL)/BST (1 QL)/MnBST (1 SL)/BST (1 QL) with various Sb contents, $x$. a,** Magnetic field dependences of Hall resistivity ($\rho_{yx}$) at Sb content of $x$ = 0.20 - 0.60. Inset of **a**: Enlarged image of **a** at the low magnetic fields at $x$ = 0.20, 0.50 and 0.60. **b-d,** Sb contents dependences of 2D carrier type/density ($n_{2D}$) (**b**), $|\rho_{yx}|$ at zero magnetic field (**c**), and magnitudes of anomalous and topological Hall effects (**d**). Both $|\rho_A^{sat}|$ and $|\rho_T^{peak}|$ are enhanced when the $E_F$ locates closer to the CNP. **e,** Schematic picture of the relation between the position of the $E_F$ and the stability of skyrmions (Sk). When the $E_F$ locates near the edge of the gapped Dirac cone, skyrmions are stable.

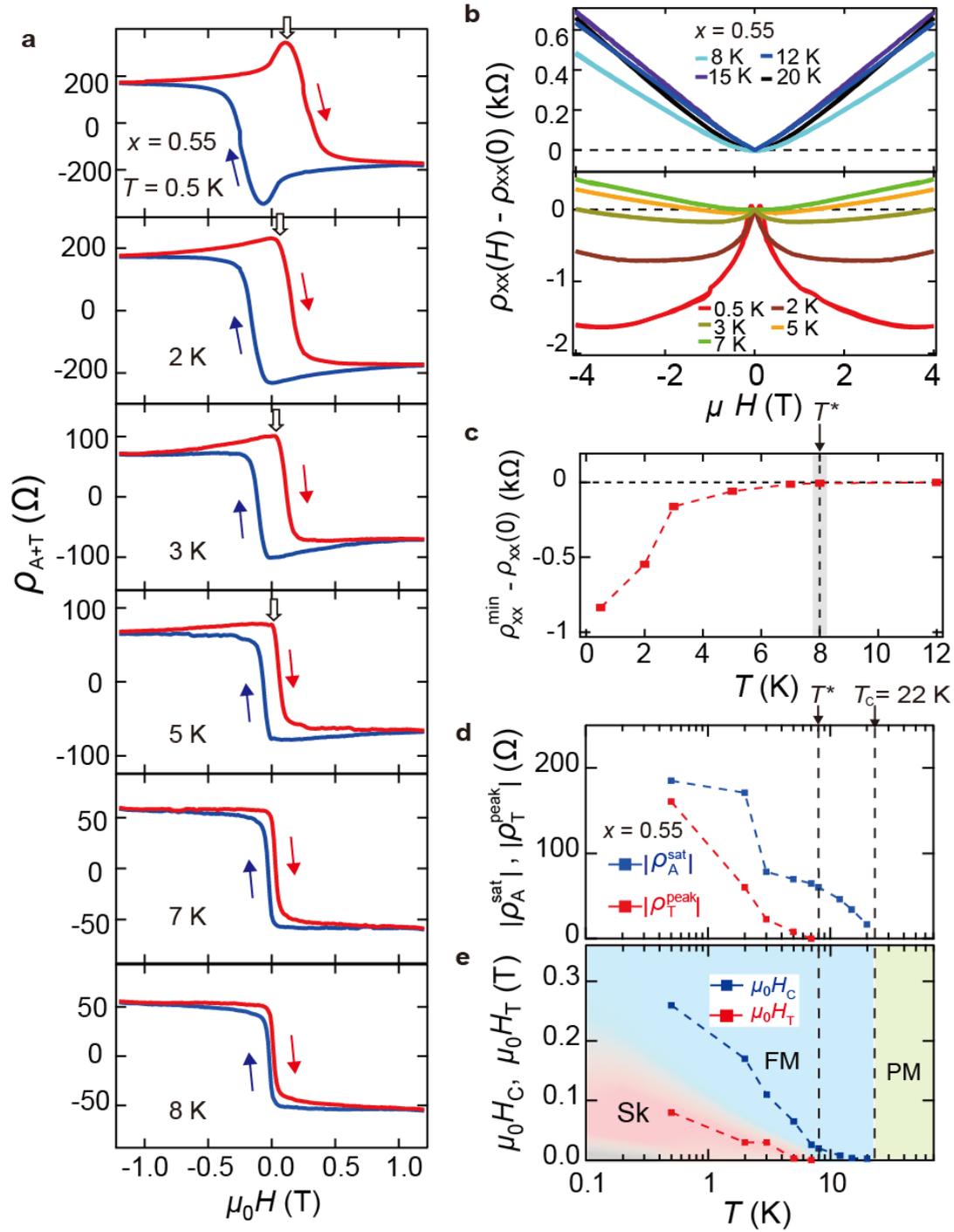

**Figure 3 | AHE, THE and MR of BST/MnBST/BST/MnBST sandwich structure with *x* = 0.55.**
**a,** Magnetic field dependences of the Hall resistivity ($\rho_{A+T}$) where the ordinary Hall-effect component is excluded. **b,** MR (= $\rho_{xx}(H) - \rho_{xx}(0)$) at temperatures of 0.5 - 20 K. **c,** Temperature dependence of the negative MR at the low magnetic field (< 1 T), the difference between resistivity at zero field ($\rho_{xx}(0)$) and minimum resistivity ($\rho_{xx}^{min}$), which almost disappears above $T = T^* \sim 8$ K. **d** and **e,** Temperature dependences of (**d**) $|\rho_A^{sat}|$ and $|\rho_T^{peak}|$, (**e**) the $H_C$, and the field at the hump $H_T$ at $T = 0.5 - 20$ K.

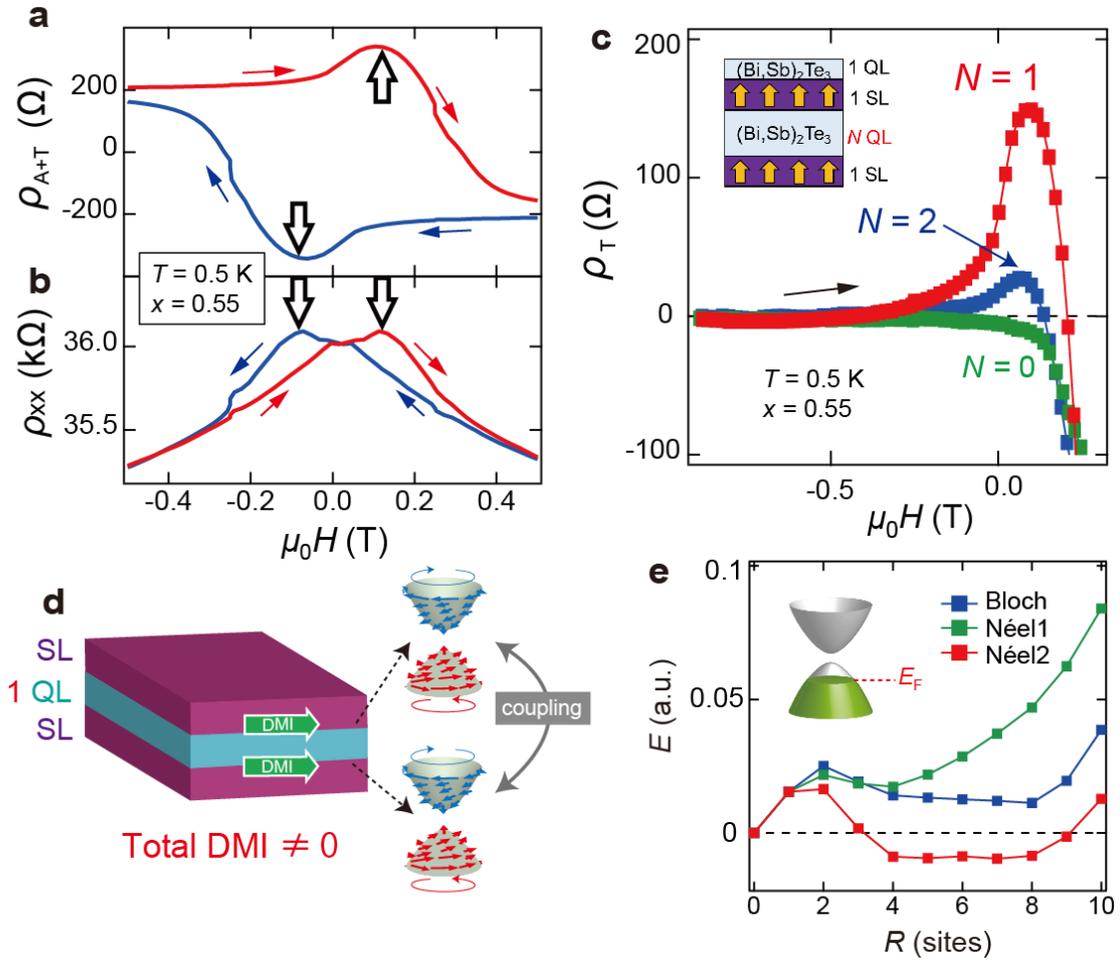

**Figure 4 | Properties of the topological Hall effect (THE) in BST/MnBST/BST/MnBST sandwich structure. a,b,** Magnetic-field dependences of (**a**) $\rho_{A+T}$ and (**b**) MR at 0.5 K. The peaks indicated by hollow arrows in (a) correspond to those at peaks in (b). **c,** Magnetic-field dependences of the THE component, $\rho_T$, which is estimated by subtraction of $\rho_A^{sat}$ from $\rho_{A+T}$, at samples with the spacer BST thicknesses of 0, 1, and 2 QL. **d,** Schematics of the coupling of surface states between top and bottom with the same spin chirality in 1 QL-BST sandwiched by two MnBST-SLs, leading to the nonzero DMI inducing skyrmions. **e,** Calculations of the energetic stability of three types of skyrmions in the BST/MnBST/BST/MnBST system. The $E_F$ is slightly below the magnetic gap of the Dirac cone. $R$ corresponds to the size of a skyrmion and "sites" indicates the in-plane length unit. The Néel2-type skyrmion is most stable (red curve).

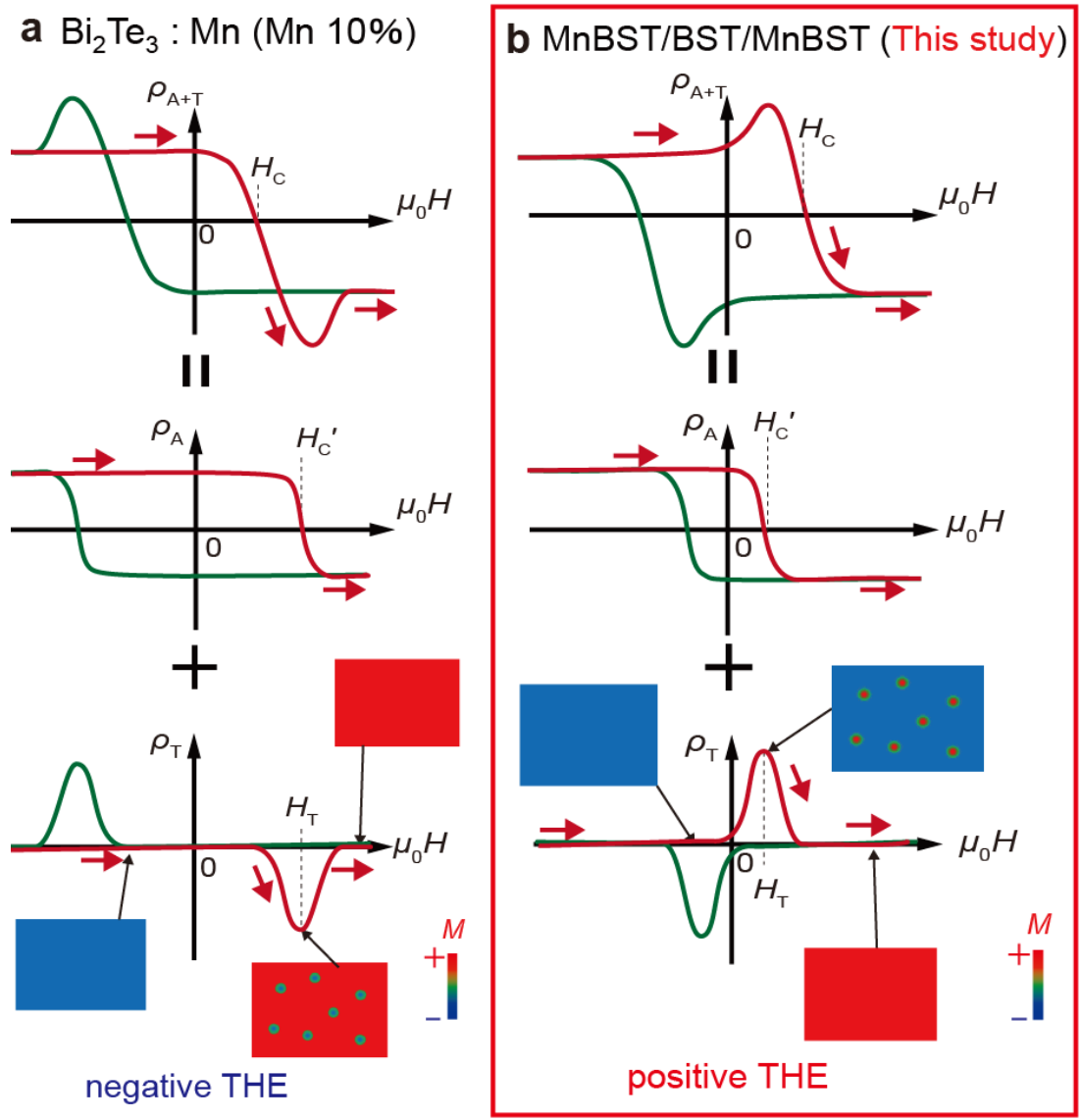

**Figure 5 | Comparison of magnetizatoin reversal between the previous Mn-doped $Bi_2Te_3$ and our MnBST/BST/MnBST. a,b,** Schematics of AHE, THE, and their sum properties in **a,** the reported system $(Mn_{0.1}Bi_{0.9})_2Te_3$ at $T = 1.5$ K[10] and **b,** the present system MnBST/BST/MnBST at $T = 0.5$ K. When the magnetic field is swept from the negative to the positive (along the red arrows), the opposite sign of the THE is observed between (**a**) Mn-doped system and (**b**) this study, indicating that the timing of generation of skyrmions is different from each other. Whereas in the Mn-doped case, skyrmions are reversed after the reversal of the matrix magnetization, in our case, they are reversed before the reversal of the matrix magnetization. Around $H_c'$, density of skyrmions and the carrier scattering reach maximum. In our case, skyrmions are generated at the lower magnetic field than the previous ones; $H_T$ in MnBST (~ 0.1 T at 0.5 K) is much smaller than that in the Mn-doped system (~ 1 T at 1.5 K).

**Supplementary Information**

**Soft-magnetic skyrmions induced by surface-state coupling in an intrinsic ferromagnetic topological insulator sandwich structure**

Takuya Takashiro[1], Ryota Akiyama[1], Ivan A. Kibirev[2], Andrey V. Matetskiy[2],

Ryosuke Nakanishi[1], Shunsuke Sato[1], Takuro Fukasawa[3], Taisuke Sasaki[4],

Haruko Toyama[1], Kota L. Hiwatari[1], Andrey V. Zotov[2], Alexander A. Saranin[2],

Toru Hirahara[3] and Shuji Hasegawa[1]

[1]*Department of Physics, The University of Tokyo, Bunkyo, Toyko 113-0033, Japan*

[2]*Institute of Automation and Control Processes, 690041, Vladivostok, Russia*

[3]*Department of Physics, Tokyo Institute of Technology, Tokyo 152-8551, Japan*

[4]*National Institute for Materials Science (NIMS), 1-2-1 Sengen, Tsukuba, Ibaraki 305-0047, Japan*

Corresponding author: akiyama@surface.phys.s.u-tokyo.ac.jp (R.A.)

**Contents:**

# I. Detailed observation of cross-sectional structure in the magnetic sandwich TI

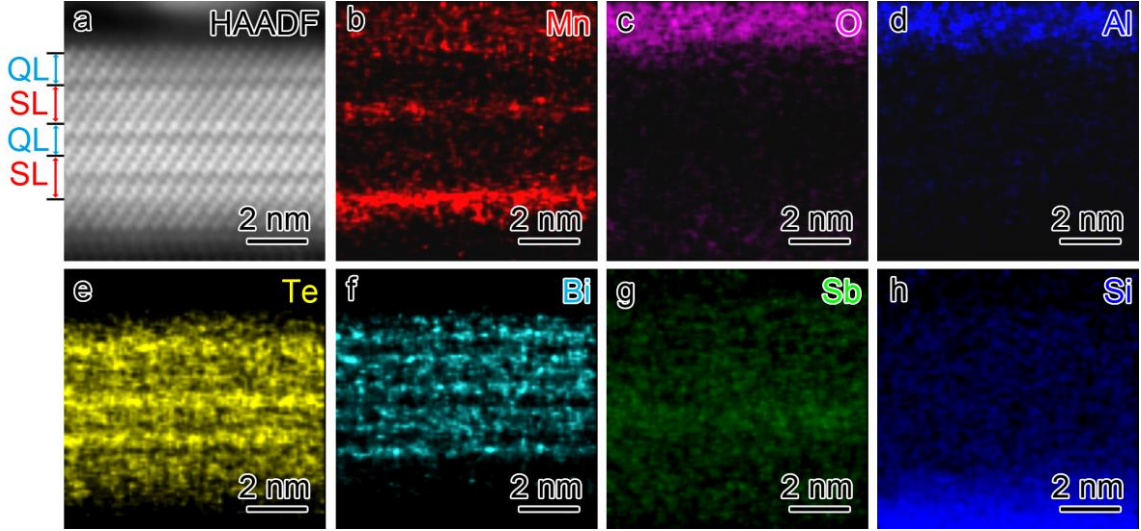

**Figure S1 | high-angle annular dark field STEM (HAADF-STEM) and EDS results. a,** Cross-sectional HAADF-STEM image of the ferromagnetic sandwich topological insulator structure. **b-h,** Distribution maps of elements, Mn, O, Al, Te, Bi, Sb, and Si in the sample as shown in **a**.

**Figures S1a through h** display high-angle annular dark field scanning transmission electron microscope (HAADF-STEM) image and energy dispersive X-ray spectroscopy (EDX) elemental maps for an $Al_2O_3$-capped QL/SL/QL/SL sandwich structure composed by $(Bi_{1-x}Sb_x)_2Te_3$ (BST) and $Mn(Bi_{1-x}Sb_x)_2Te_4$ (MnBST) on the Si substrate. Note that **Figures S1a and b** were shown in **Figure 1f** in the main text. It should be noted that Bi, Sb and Te are distributed into the QL/SL/QL/SL layer while Mn is into SLs and O and Al are located in the topmost capping layer.

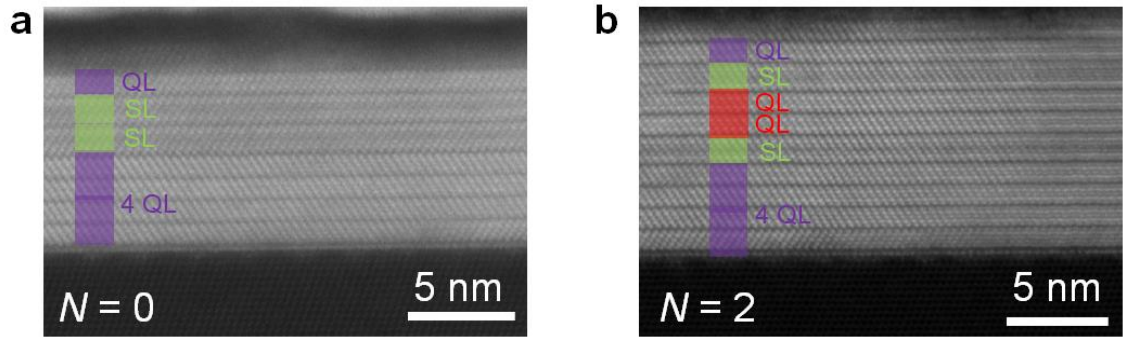

**Figure S2 | Cross-sectional HAADF-STEM images with different spacer-thickness ($N$ QL) between two SLs. a** and **b**, SLs sandwich the spacer of $N = 0$ and 2, respectively.

**Figures S2 a** and **b** show HAADF-STEM images of QL/SL/SL ($N = 0$) and QL/SL/2 QL/SL ($N = 2$), respectively, as introduced in **Figure 4c** in the main text. The sandwich structure of SL/$N$-QL/SL forms on nonmagnetic topological insulator layers (4 QL). Topological Hall effect (THE) induced by the skyrmions is related to the coupling between top and bottom surface states on MnBST layers and the spacer-thickness dependence of THE as shown in **Figure 4c** in the main text is contributed from the sandwich structures above the 4 QL.

**II. Magnetic property of the MnBST/BST/MnBST sandwich structure**

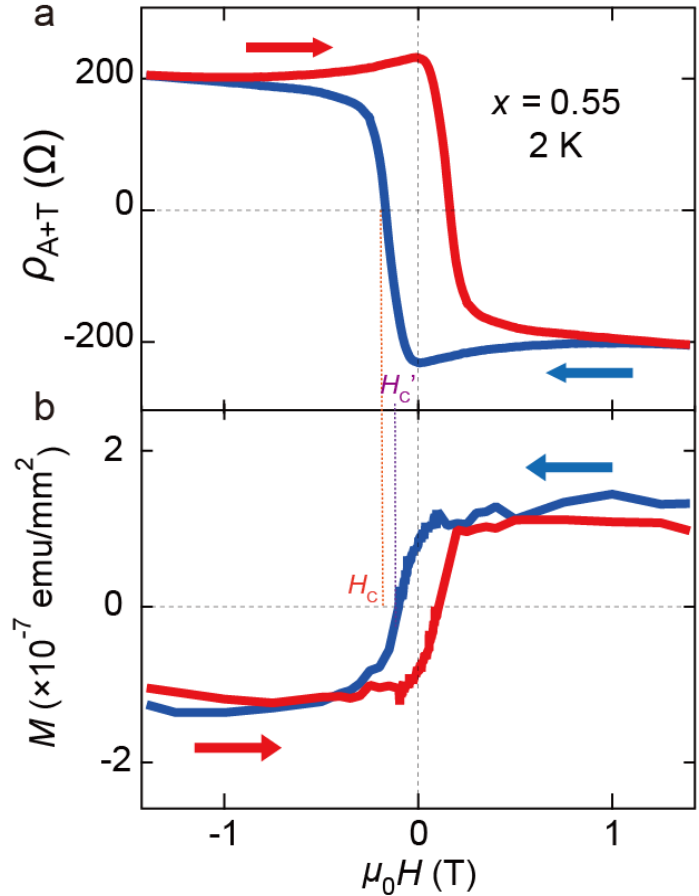

**Figure S3 | Comparison between $\rho_{A+T}$-$H$ and $M$-$H$ hysteresis loops in the ferromagnetic topological insulator sandwich structure (QL//SL/QL/SL sample) at 2 K. a,** Magnetic field dependence of the Hall resistivity excluding the ordinary Hall effect. **b,** Magnetic field dependence of magnetization measured by SQUID. $H_C$ and $H_C$' are not coincident because $H_C$ is affected by not only AHE but also THE.

Figures S3a and S3b display the anomalous and topological Hall property of QL/SL/QL/SL sample in the electrical transport measurement at $T$ = 2 K, and the magnetic property of QL/SL/QL/SL sample measured by a superconducting quantum interference device (SQUID) based magnetometer at $T$ = 2 K, respectively. The $M$-$H$ curve by SQUID indicates that the top and bottom MnBST layers have the same ferromagnetic properties because there is no additional step near the coercivity of the hysteresis loop. It should be noted that the $M$-$H$ coercivity ($H_C'$ ~ 0.09 T) is smaller than the $\rho_{A+T}$-$H$ coercivity ($H_C$ ~ 0.16 T) because $H_C$ contains the effect of the THE. The similar tendency was reported in the previous THE-studies[36,37]. The SQUID measurement gives the net-magnetization including matrix and skyrmions so that the number of generated skyrmions reaches maximum around $H = H_C$.

Estimated from the saturated magnetization in the $M$-$H$ curve, the effective magnetic moment of Mn at 2 K is ~ $1.1\mu_B$ where $\mu_B$ represents the Bohr magneton. This value is comparable to those of the multi-layered $MnBi_2Te_4$ ($1.14\mu_B$ at 3 K)[38] and the Mn-doped $Bi_2Te_3$ ($1.5\mu_B$ at 1.8 K)[39], which is much smaller than $5\mu_B$ expected for $Mn^{2+}$ ions. This supports that the magnetic order in the MnBST/BST/MnBST sandwich structure can be explained by the mechanism of itinerant ferromagnetic systems, which is associated by the RKKY mechanism as mentioned in the main text.

## III. Symmetrizing and anti-symmetrizing procedures in the magnetic field dependence of electrical transport

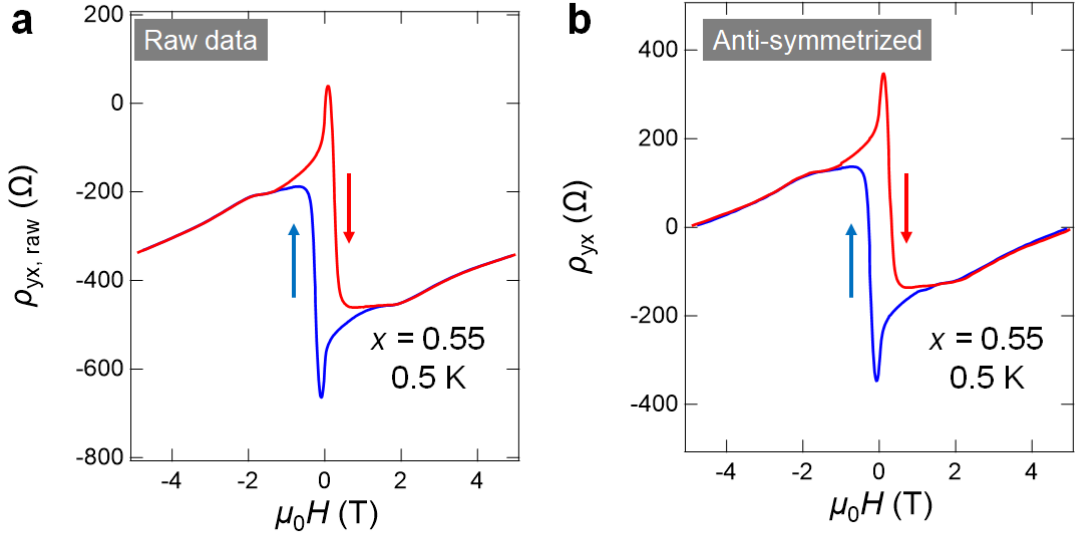

**Figure S4 | Anti-symmetrizing procedure for Hall resistivity. a,** the raw data in Hall measurement in the QL/SL/QL/SL sample with $x$ = 0.55 at 0.5 K. **b,** the anti-symmetrized data of **a**.

All the raw data of MR (Hall) measurements has some offset and is unexpectedly affected by the Hall (MR) properties because of the possible slight misalignment of the contact electrodes. To exclude these effects, the Hall and MR raw data were anti-symmetrized or symmetrized as functions of magnetic field, respectively;

$$\rho_{xx}(B,M) = \frac{\rho_{xx,raw}(B,M) + \rho_{xx,raw}(-B,-M)}{2}, \quad (1)$$

$$\rho_{yx}(B,M) = \frac{\rho_{yx,raw}(B,M) - \rho_{yx,raw}(-B,-M)}{2}. \quad (2)$$

**Figure S4a** shows the raw data of Hall resistivity the QL/SL/QL/SL sample with $x$

= 0.55 at $T$ = 0.5 K, for which the above anti-symmetrizing was performed and the resultant data as shown in **Figure S4b** was obtained. Such symmetrizing or anti-symmetrizing procedures was applied for all the data presented in this study with careful treatments. Note that these procedures do not affect the intrinsic behavior of the data in this paper.

# IV. Fermi level shift in $(Bi_{1-x}Sb_x)_2Te_3$ with co-depositing Mn and Te

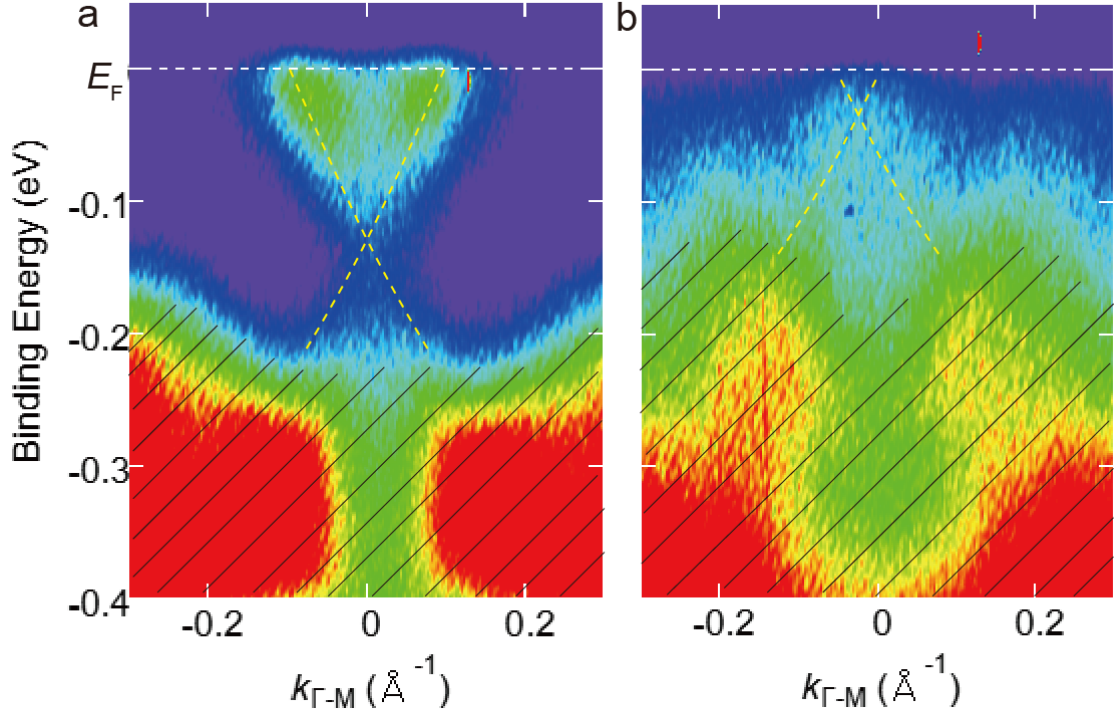

**Figure S5 | Fermi level ($E_F$) shift in $(Bi_{0.11}Sb_{0.89})_2Te_3$ (BST) by co-deposition of Mn and Te.** ARPES images of **a,** pristine BST and **b,** Mn$(Bi_{0.11}Sb_{0.89})_2Te_4$/BST (MnBST/BST) at $T$ = 79 K. The $E_F$ (white dash line) shifts around 90 meV downward against the Dirac cone by co-deposition of Mn and Te in addition to the modulation of the bulk valence band modulation.

**Figures S5 a** and **b** show the band structures of the pristine BST and MnBST/BST with $x$ = 0.89 by angle-resolved photoemission spectroscopy (ARPES), respectively. Both have the Dirac cones indicated by the yellow dash lines in **Figure S5**. In the pristine BST, the $E_F$ is ~0.13 eV above the Dirac point, and with depositing of Mn and Te, the $E_F$ shifts ~ 90 meV downward for the Dirac point because the hole carrier is doped. Due to this $E_F$-shift, the charge neutral point ($x$ = 0.50-0.55) in our samples shown in **Figure 2** in the main text is different from that in the previously reported pristine BST ($x$ = 0.94-0.96)[40].

## V. Temperature dependences of magnetoresistivity

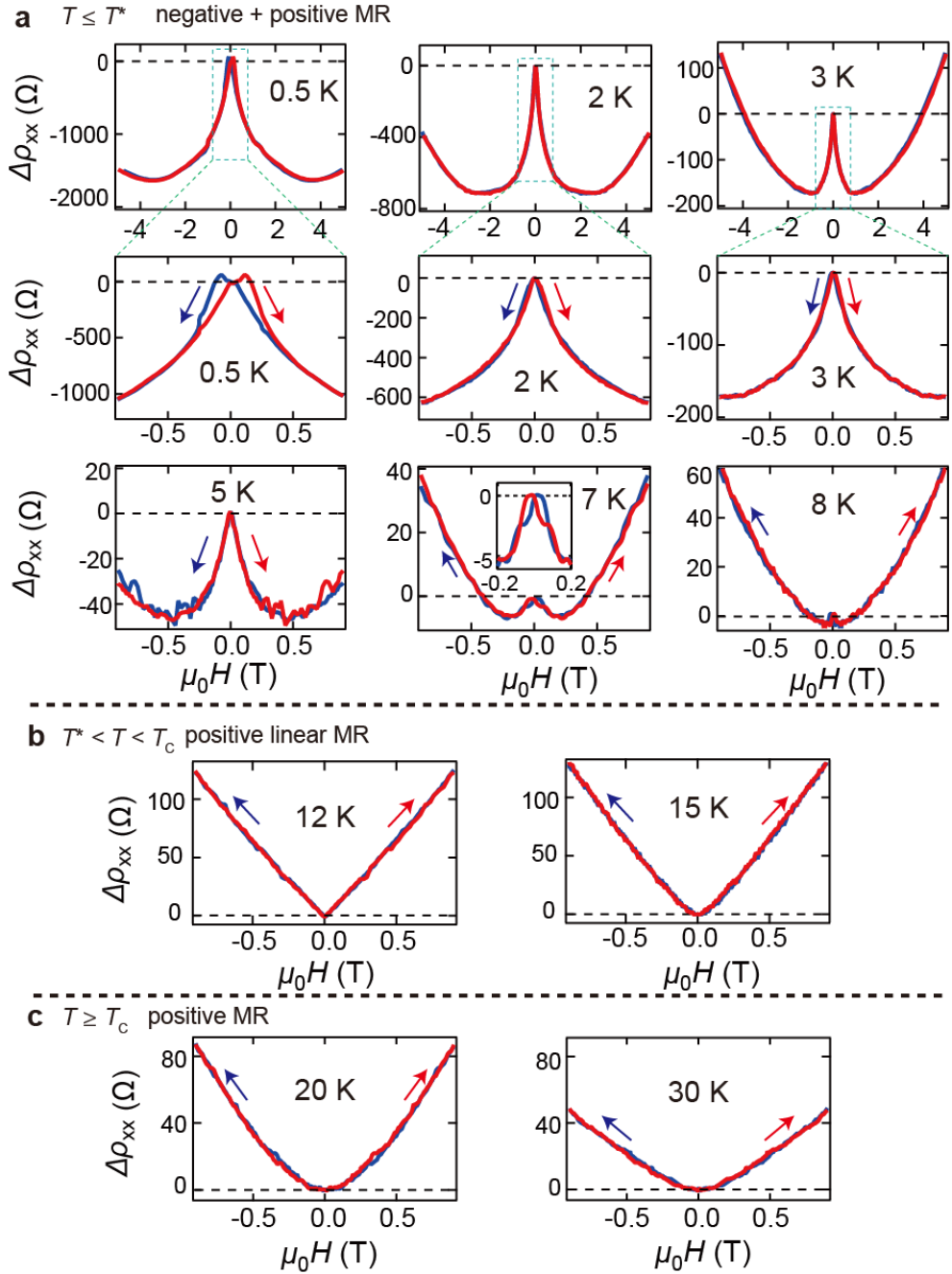

**Figure S6 | MR curves in the QL/SL/QL/SL sample with $x = 0.55$ at $T = 0.5 – 30$ K. a,** MR curves at 0.5, 2, 3, 5, 7 and 8 K, $T \leq T^*$ (~ 8 K). The red and blue curves are the sweep direction of the magnetic field. The MR curves show positive type and the cusp-like shapes at the low magnetic field. At 8 K around $T = T^*$, the negative MR becomes quite small. **b,** At 12 and 15 K at $T > T^*$, the MR is only positive type and its shape is linear, which seems a linear MR (LMR). **d,** At 20 and 30 K, $T \geq T_C$, the MR becomes positive and non-linear again.

**Figure S6** displays the MR ($\Delta\rho_{xx}$ defined as $\rho_{xx}(H) - \rho_{xx}(H = 0)$) at $T$ = 0.5 – 30 K in the QL/SL/QL/SL sample which shows the clear topological Hall effect (enlarged graphs of the MR in **Figure 3b** in the main text). At $T < T^*$ ($T^* \sim$ 8 K; derived in **Figure 3c**), the MR shows the negative cusp-like curves at the low magnetic field in addition to the positive type at the high magnetic field, and especially the hysteresys curve is clearly seen at 0.5 K as shown in **Figure S6a**. With increasing temperature, the negative cusp becomes smaller and it becomes almost negligible at $T = T^*$ ($\sim$ 8 K). When temperature is over $T^*$ the MR shows only the positive type (**Figure S6b**). At higher temperature of 12 and 15 K, the shape of the MR changes to linear, which seems to linear MR (LMR) as previously reported in such as a TI or Bi thin films[41,42]. Lastly at 20 and 30 K, $T \geq T_C$, the MR becomes positive type and non-linear again. This would be because LMR is quantum transport effect and in this temperature range quantum coherence length is too short to observe. As the THE hump-like curves disappear around $T$ = 7 K and this is close to $T^*$, the negative MR can be considered as the scattering induced by skyrmions[43].

# VI. Evaluation of the Curie temperature by Arrott plots

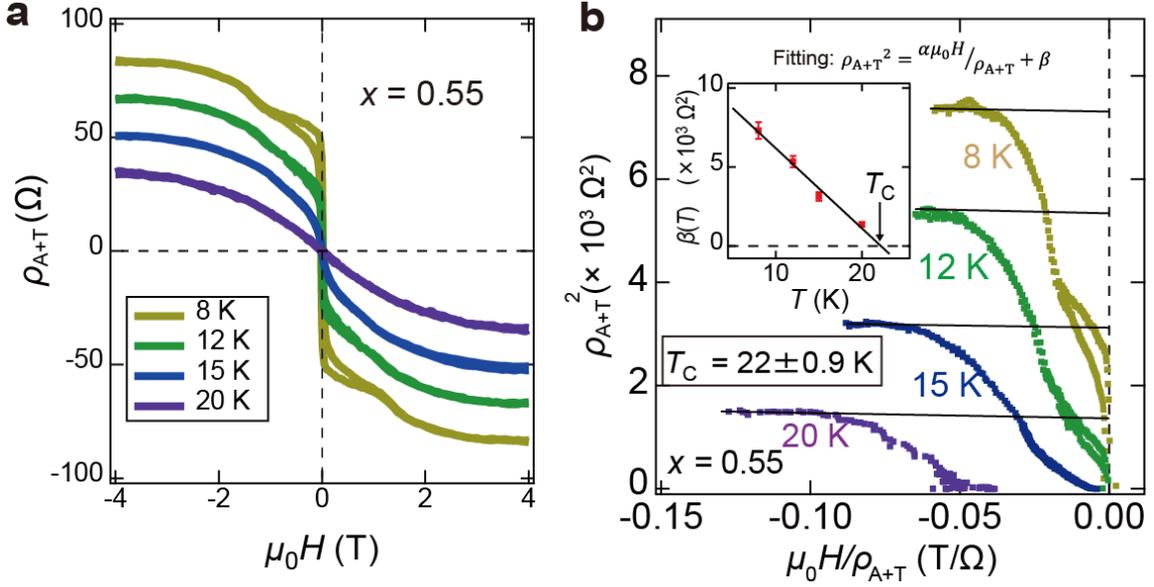

**Figure S7 | Evaluation of the Curie temperature. a,** Magnetic field dependences of anomalous Hall resistivity $\rho_{A+T}$ at temperatures of 8 – 20 K, where the THE component is negligible. **b,** Arrott plots derived from $\rho_{A+T}$-$\mu_0 H$ ranging at $T$ = 8 - 20 K. The inset shows $T_C$ = 22 K.

With the obtained AHE displayed in **Figure S7a**, we evaluated the Curie temperature ($T_C$) by Arrott plots as shown in **Figure S7b**. The extrapolating fitting lines in **Figure S7b** are the form of $\rho_{A+T}^2 = \alpha\mu_0 H/\rho_{A+T} + \beta$, where $\alpha$ and $\beta$ are the fitting parameters. Because below 7 K the significant THE component appears in the Hall resistivity, we adopted the Hall resistance above 8 K for Arrott plots[44]. The inset of **Figure S7b** indicates $T_C$ = 22 ± 0.9 K, which is close to the Curie temperature of 20 K reported in the MBE fabricated multi-layered AFM-MnBi$_2$Te$_4$[38].

# VII. Spacer-thickness dependences of ordinary, anomalous and topological Hall effects

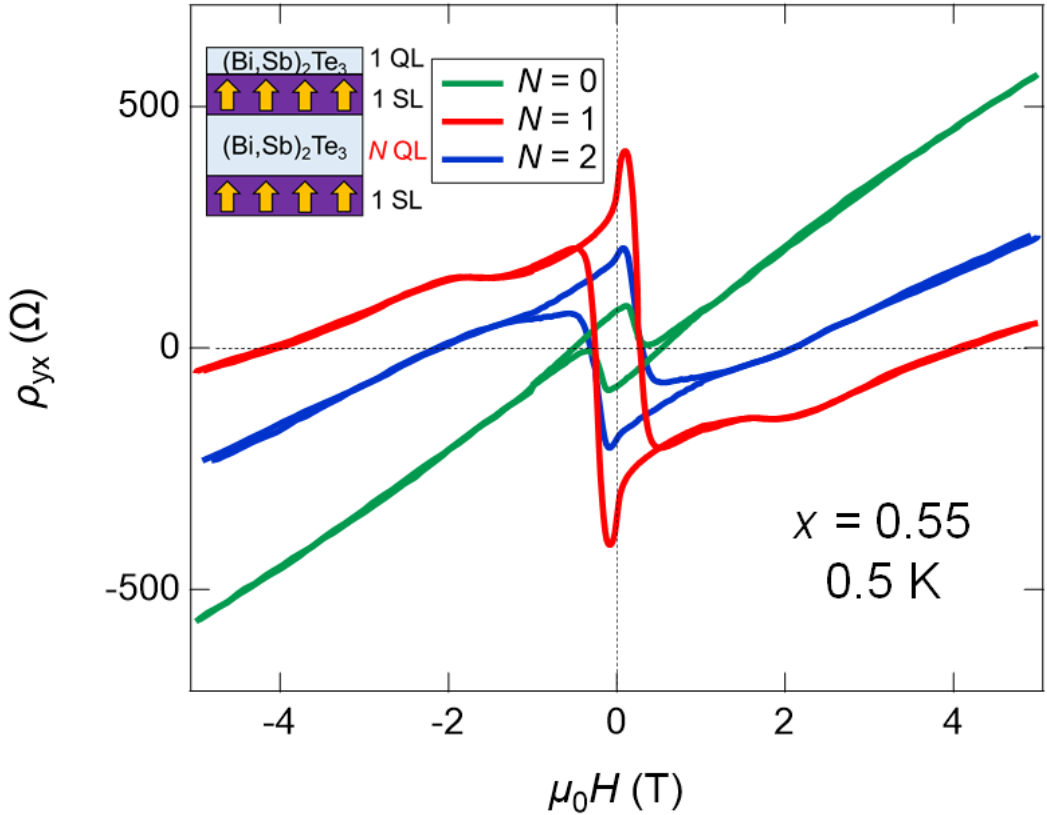

**Figure S8 | Spacer-thickness dependences of Hall effects.** Magnetic field dependences of $\rho_{yx}$ in three samples with the thickness of spacer BST, 0, 1 and 2 QL, with $x = 0.55$ at the magnetic field ranging from -5 to 5 T at $T = 0.5$ K.

**Figure S8** shows the spacer-thickness dependences of the Hall resistivity $\rho_{yx}$ in the QL/SL/$N$QL/SL ($N$ = 0, 1 and 2). The topological Hall effect can be derived by excluding the ordinary Hall effect and AHE as shown in **Figure 4c** in the main text. The carrier type is $p$-type in all samples and the carrier densities with $N$ = 0, 1 and 2 are 5.5, 10.3 and 7.8

×10$^{12}$ cm$^{-2}$, respectively, which are estimated from the linear components by the ordinary Hall effect at the high magnetic fields (> 4 T). The AHE curve with $N = 1$ is the largest, and this result is the same tendency to the spacer-thickness dependence of the THE shown in the main text.

The hysteresis loops in **Figure S8** indicate that all samples are ferromagnetic regardless of spacer-thickness. To confirm the magnetic interaction in detail, we measured the Hall properties of these samples in the higher magnetic field range from -14 to 14 T in **Figure S9**. If there is an interlayer antiferromagnetic coupling in our system as reported in the multi-layered MnBi$_2$Te$_4$[38], a spin-flip process should appear at the high magnetic field region. However, as shown in **Figure S9**, every sample show no step-like behavior by spin-flip, and our system has not antiferromagnetic but ferromagnetic interaction.

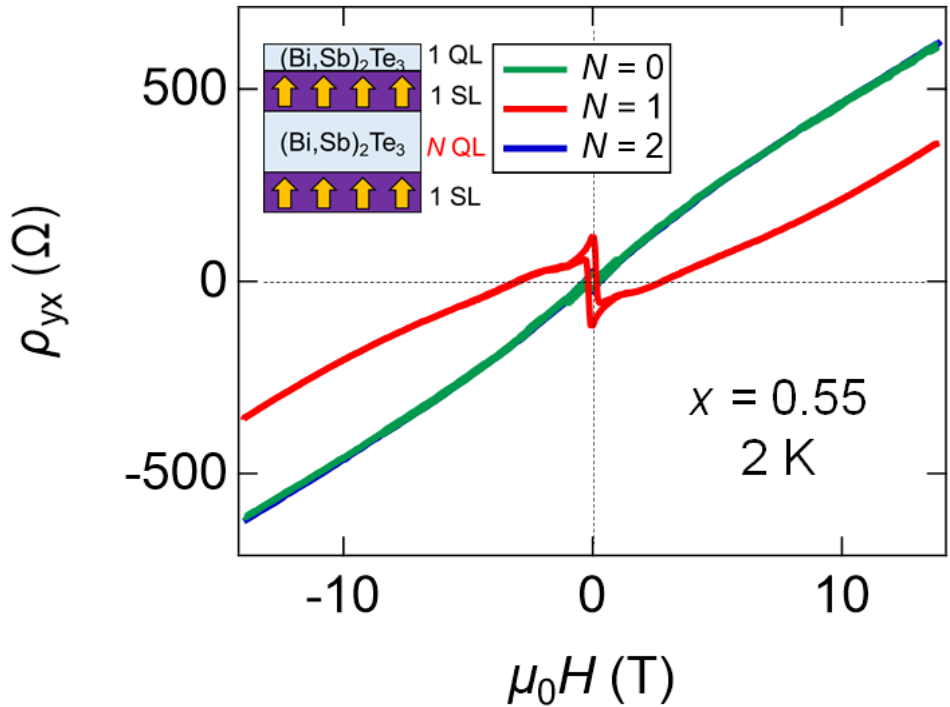

**Figure S9 | Spacer-thickness dependence of Hall effects at the higher magnetic field.** Magnetic field dependences of $\rho_{yx}$ in three samples with the thickness of the BST spacer, 0, 1 and 2 QL, with $x = 0.55$ at the magnetic field ranging from -14 to 14 T at $T = 2$ K.

**VIII. Numerical calculations of the energy stability of three-type skyrmions**

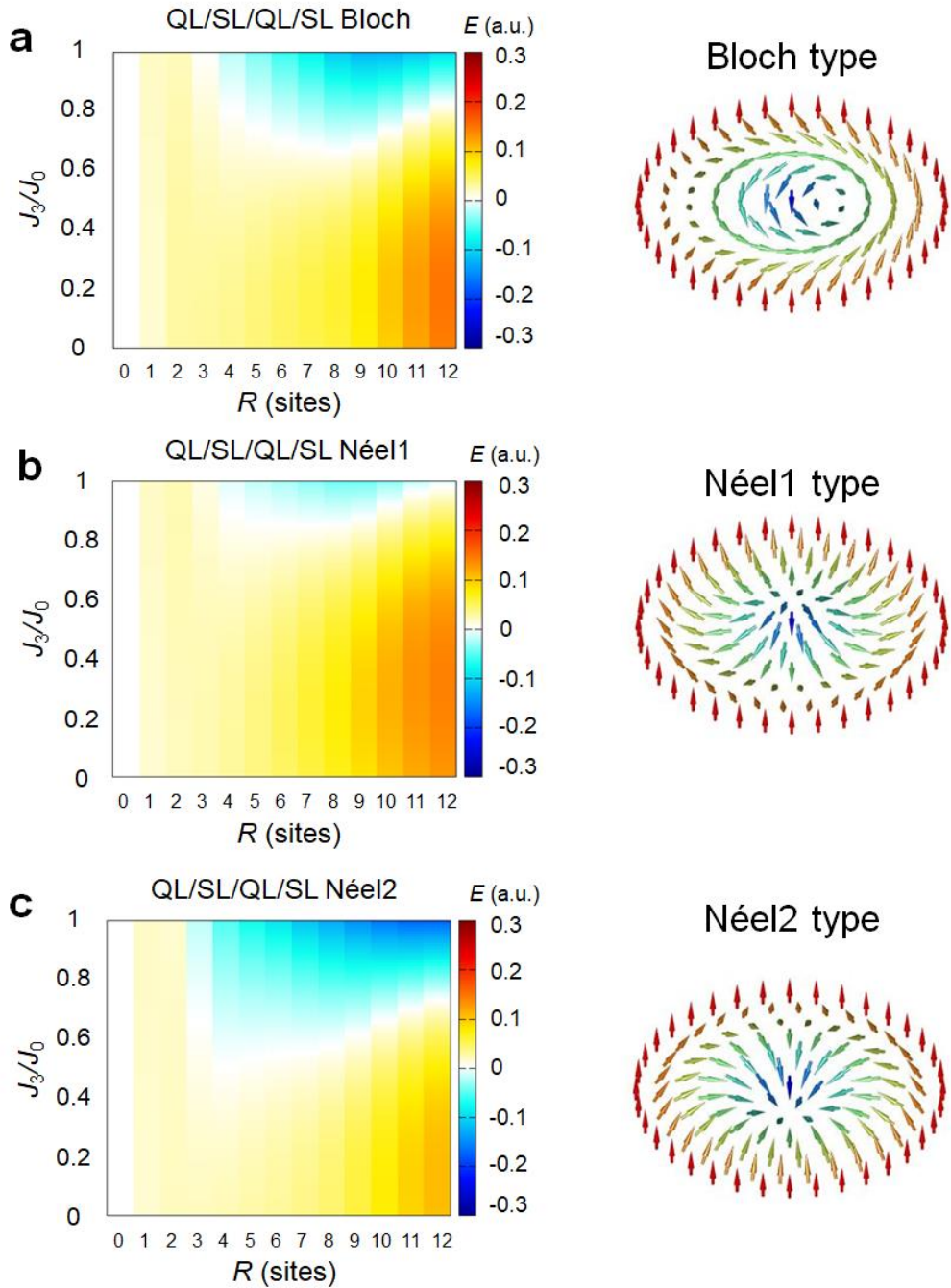

**Figure S10 | Numerical calculations of the energy stability of three types. a,** Bloch, **b,** Néel1 and **c,** Néel 2 types. The most stable configuration of the skyrmion state is Néel 2 type because the negative energy area of Néel 2 type in three color plots is the largest.

As explained in the main text and **Methods**, we numerically calculated the energy stability of three types of skyrmions (Bloch, Néel 1 and Néel 2 types) in the BST/MnBST/BST/MnBST structure. **Figure S10** shows color maps of the energy stability as functions of $J_3/J_0$ and $R$ for three types of skyrmions where $J_0$ and $J_3$ represent the symmetric, asymmetric parts of the magnetic exchange couplings for the two orbitals which belong to the conduction and valence band each other, respectively. $R$ is the skyrmion size. Here, $J_3$ varies from 0 to $J_0$ and the other parameters are same values written in **Methods**. These results mean that if $J_3$ is around zero, the energy of the ferromagnetic order ($R = 0$) is lower than those of all types of skyrmions ($R \neq 0$), indicating that the ferromagnetic order is more stable than skyrmion states. If $J_3 > 0.6J_0$, skyrmion states become more stable. In our case, the most stable configuration of skyrmions is Néel 2 as written in the main text since it is stable for the broader ranges of $J_3/J_0$ and $R$, which is the same tendency as previous numerical calculations[36,45].

**IX. Discussion about the possibility of "mimic topological Hall effect"**

K. M. Fijalkowski *et al.* have recently indicated that the characteristic hump-like curves observed in a magnetically-doped TI/nonmagnetic TI heterostructure in the Hall measurement can be due to not the skyrmion-induced THE[46] but the sum of curves of the opposite sign AHE, which result from two FM orders in the bulk and at the surface. However, the hump-like curves in our system cannot be explained based on above assumption of the two coexisting different FM orders from the following points: (i) Our system can exclude bulk FM orders. (ii) THE has the spacer-thickness dependence. (iii) the *M-H* curve by SQUID shows no step-behavior and cusp.

Regarding (i), our system has the two-dimensional single Mn layer possessing the FM order on the top and bottom of a TI, and thus the bulk FM component is not incorporated and there is no another FM order than the intrinsic one in SLs.

Next (ii), if the hump-like curves result from coexisting of the two AHEs having an opposite sign, it should neither disappear nor be weakened by changing the thickness of the BST spacer. However, in our case the hump-like curves significantly change as shown in **Figure 4c** in the main text.

Finally about (iii), if our system had the two FM components, the *M-H* curves would show step-behaviors corresponding to the magnetization reversal of the two FM components in addition to the cusp which is generated by sum of the two AHEs. However, our SQUID result (**Figure S3b**) shows no such a step and cusp behavior.

These facts confirm that our hump-like curves in the Hall effect are formed by the skyrmion-induced THE.

**X. Estimation of skyrmion density and length scale**

In the magnetic skyrmion systems in chiral magnets and FMTI, the magnitude of THE allows us to estimate the skyrmion density and length scales in a semi-quantitative manner[36,37,45]. Skyrmions possess the emergent magnetic field ($H_{em}$) proportional to the skyrmion density ($n_{Sk}$) and the flux quantum ($\phi_0 = h/e$) as

$$\mu_0 H_{em} = -n_{Sk}\phi_0 \equiv B_{em} \quad (3)$$

leading to the induction of THE expressed by

$$\rho_{THE} = P R_0 n_{Sk} \phi_0 = P R_0 B_{em} \quad (4)$$

where $P$ is spin polarization of conduction electrons in the film and $R_0$ is the ordinary Hall coefficient. As the spin polarization of V, Cr and Mn doped $Bi_2Te_3$ is estimated as 50% in previous works[45,47,48], we here set the $P$ of BST/MnBST/BST/MnBST to be in the same order of magnitude. The $R_0$ is derived from the slope of the linear part of Hall resistivity. Using the topological Hall resistivity $\rho_T$ defined in the main text, the skyrmion density in the $N = 1$ sample with $x = 0.55$ is estimated to be $2.3 \times 10^{15}$ m$^{-2}$ at $T = 0.5$ K, and $1.3 \times 10^{15}$ m$^{-2}$ at 2 K. The length scale of a single skyrmion ($r_{Sk}$) can be estimated as $(n_{Sk})^{-1/2}$ to be ~ 21 nm at 0.5 K and ~ 28 nm at 2 K. Such values are comparable to the typical length scale of Dzyaloshinskii-Moriya interaction (DMI) induced skyrmions ranging from 1 to 100 nm[49]. The skyrmion size is determined by the ratio of the strength of DMI (*D*) and ferromagnetic exchange interaction ($J$), that is, nearly proportional to $J/D$ [50]. The previously observed skyrmion in Mn-doped $Bi_2Te_3$ has $r_{Sk}$ ~ 30 nm at 1.5 K, which is almost the same as ours, indicating that there is no large difference in $J/D$ between two studies[45]. As explained in the main text, $J$ in our system is suggested to be larger than that in the Mn-doped system, and considering above we could say that the DMI in our system is also larger.